\documentclass[10pt,draft]{dis03}
\usepackage{epsf,amsmath}
\usepackage{epsfig}

\newcommand{\as}{\alpha_s}
\newcommand{\ee}{e^+e^-}
\newcommand{\cO}[1]{\mathcal {O}\left(#1\right)}
\begin{document}

\title{JET OBSERVABLES IN HADRONIC DIJET PRODUCTION}
  
\author{ANDREA BANFI \\
NIKHEF Theory group\\
P.O. Box 41882 1009 DB Amsterdam, the Netherlands\\
E-mail: andrea.banfi@nikhef.nl }

\maketitle

\begin{abstract}
\noindent We present a numerical program, CAESAR, 
that allows us to resum large logarithmic contributions to jet
observables in a fully automated way. As an application we obtain the
first next-to-leading logarithmic distributions for event shapes in
hadronic dijet production.
\end{abstract}

\section{Why hadronic collisions?} 
Hadronic collisions constitute one of the most exciting environments
for high energy physicists, both theorists and experimentalists. Due
to the large centre-of-mass energies that can be reached, they are
ideal for the search of new particles. Moreover, they are incredibly
rich from the point of view of QCD. In particular, measures of
final-state energy flow are important for various aspects.

Besides for the `standard' measurements of the coupling constant $\as$
\cite{Bethke} and the colour factors \cite{SU3}, final state
observables are particularly suited for investigations of the yet
poorly known infrared domain of QCD, since they are affected by large
non-perturbative (NP) contributions , originating both from
hadronisation corrections \cite{DMW} and from the so-called `soft
underlying event' \cite{MW}. 

Event shapes and jet rates (both referred to as `jet observables') are
among the most studied of such measures.  The study of jet observables in
hadronic dijet production is at the very beginning.  Theoretically
only fixed order perturbative (PT) calculations are available
\cite{NLOJET}, and there exist experimental data for just one event
shape \cite{D0}.  Both studies show clearly that in the region where
the observables are small, large infrared and collinear logarithms
arise that need be resummed to obtain a sensible answer.
The involved kinematics of the process makes analytic calculations
really cumbersome, so that such observables are the ideal testing
ground for an automated resummation approach such as the one provided
by the program CAESAR \cite{BSZ}.  The program is based on a master
formula whose input is determined numerically in a preliminary stage.
The user needs only to provide a subroutine that
computes the observable given a set of four-momenta. The program then
returns the observable's resummed distribution $\Sigma(v)$ (the
fraction of events for which the observable's value is less than $v$) at
next-to-leading logarithmic (NLL) accuracy.\footnote{We recall that
  NLL accuracy means resumming all $\exp\{\as^n \ln^{n+1} 1/v\}$ and
  $\exp\{\as^n\ln^n 1/v\}$ terms in $\Sigma(v)$ \cite{CTTW}. } All the
details of the approach have been explained by Giulia Zanderighi
\cite{Giulia}.  In the following we describe through some examples
what kind of observables can be studied with CAESAR, and present some
output distributions.

\section{Observable definition}
The first observable we introduce in hadronic collisions is the
transverse thrust, which represents the analogous of the
thrust in $\ee$ in the plane orthogonal to the beam axis:
\begin{equation}
  \label{eq:thrustT}
  T_t = \max_{\vec n_t} 
  \frac{\sum_i|\vec p_{ti} \cdot \vec n_t|}{\sum_i |\vec p_{ti}| }\>,
\end{equation}
where the sum runs over all possible final state hadrons of transverse
momenta $p_{ti}$, and $\vec n_t$ is a unit transverse
vector.\footnote{Here transverse means orthogonal to the beam.}
The distribution in this observable needs resummation in the region
$\tau_t\equiv 1-T_t \ll 1$. 

It is clear that the sum in \eqref{eq:thrustT} cannot include all the
final state particles, since any actual measurement necessarily
excludes a region around the beam. Typically one measures only
particles whose rapidity $\eta$ is in the range $|\eta| < \Delta$.  In
such a situation a particular class of NLL contributions emerges, the
so-called `non-global logs' \cite{NG}, which arise whenever an observable is
sensitive to secondary particle emission only in a limited region of
the phase space. Their presence causes a loss of accuracy in NLL
predictions, since their expression is known only in the large $N_c$
limit. 

Fortunately, any observable can be made global with just small
modifications of its definition, as we show in the following two examples.
\begin{enumerate}
\item Directly global thrust $\tau_{t,g}$:
  \begin{equation}
    \label{eq:global-T}
    \tau_{t,g} = 1-\max_{\vec n_t} 
  \frac{\sum'_i|\vec p_{ti} \cdot \vec n_t|}{\sum'_i |\vec p_{ti}| }\>, 
  \end{equation}
  where the sum now runs over all hadrons with $|\eta| < \Delta$ and
  $\Delta$ is taken as large as possible. Observables of this kind are
  actually non-global, but non-global effects do not show up at NLL
  accuracy as long as $v \geq e^{-c\Delta}$, with $c$ an observable's
  dependent coefficient (in this case $c=1$) \cite{BSZ}.
\item Indirectly global thrust $\tau_{t,\Delta}$:     
  \begin{equation}
    \label{eq:recoil-T}
    \tau_{t,\Delta} = 1-\max_{\vec n_t} 
  \frac{\sum'_i|\vec p_{ti} \cdot \vec n_t|}{\sum'_i |\vec p'_{ti}| }+R_t\>,
  \qquad
  R_t=\frac{\left|\sum'_i \vec p_{ti}\right|}{\sum'_i |\vec p_{ti}| }
  \>. 
  \end{equation}
Here again the sum runs over all particles with $|\eta| < \Delta$, but
$\Delta$ can be taken of $\cO{1}$, since the recoil term $R_t$, due to
transverse momentum conservation, makes the observable sensitive also
to particles inside the beam region. 

Indirectly global observables are known to cause consistency problems
for NLL resummations \cite{numsum}.  Actually what any NLL answer
assumes is that an observable is kept small by forbidding
particle emission above a given momentum scale. In this case
$\tau_{t,\Delta}$ can be small not only because all involved momenta
are required to be small, but also because vectorial cancellations
occur among larger transverse momenta.  It also happens that while the
probability of vetoing radiation decreases with the observable's
value, that of having vectorial cancellations is independent of that
value.  Therefore, below a given $\tau_{t,c}$ the second mechanism
overcomes the first, and NLL predictions break down developing a
singularity.  However, as long as $\tau_{t,\Delta}$ gets not too close
to $\tau_{t,c}$ NLL predictions are still meaningful.  The particular
choice of the recoil term in \eqref{eq:recoil-T} ensures that
$\tau_{t,c}$ is away from the range of values of $\tau_{t,\Delta}$ that are
accessible through PT calculations.
\end{enumerate}

Analogously we define the two version of the thrust minor
\begin{equation}
  \label{eq:minor}
  T_{m,g} =  \frac{\sum'_i|\vec p_{ti}\times \vec n_t|}
  {\sum'_i|\vec p_{ti}|}\>,\> \qquad
  T_{m,\Delta} =  \frac{\sum'_i|\vec p_{ti}\times \vec n_t|}
  {\sum'_i|\vec p_{ti}|}+R_t\>,
\end{equation}
which are both measures of the energy flow out of the plane formed by
the beam and the thrust axis $\vec n_t$.

\section{Some worked out examples} 
\begin{figure}[!htb]
\vspace*{5.cm}
  \begin{center}
    \includegraphics{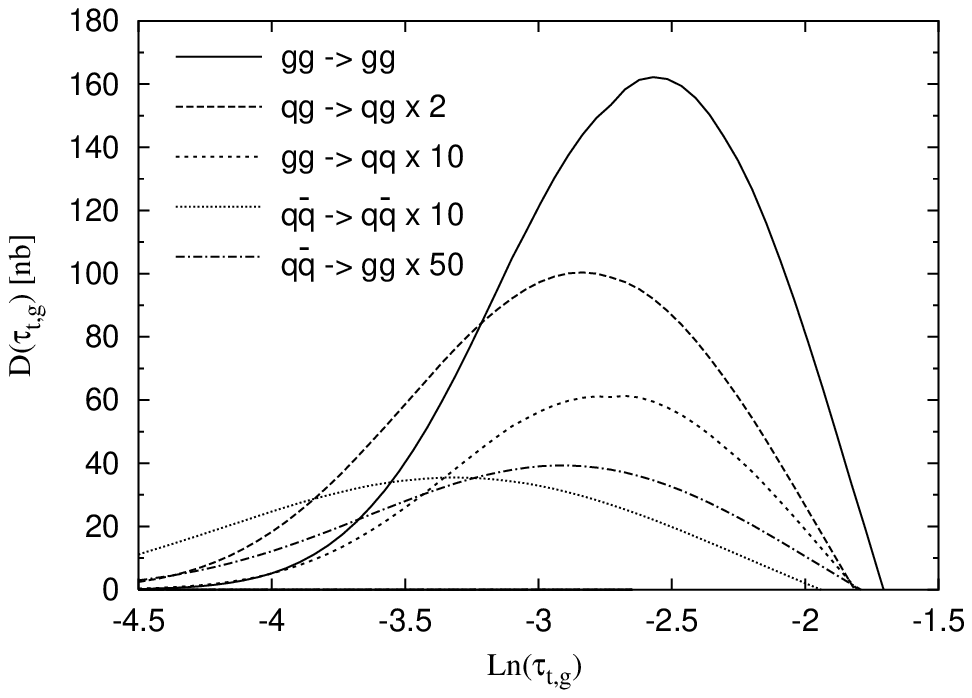}
    \includegraphics{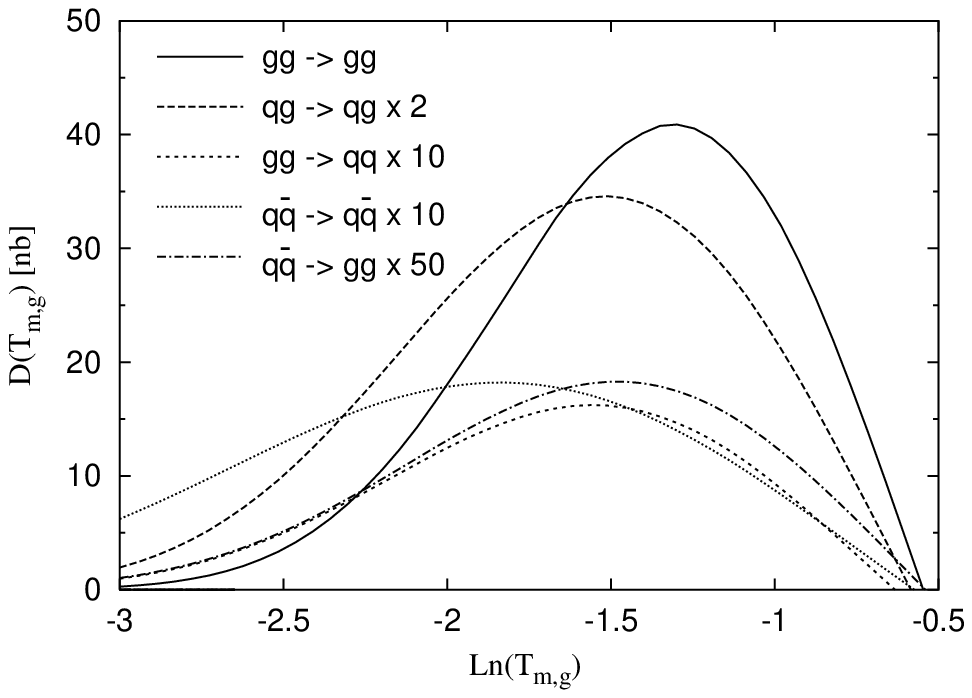}
  \caption[*]{The resummed differential distributions $D(v)\equiv d\Sigma(v)/d \ln v$ for the global transverse thrust (left) and thrust minor (right).}
  \label{fig:T-Tmin}
\end{center}
\end{figure}

We present results for the directly global thrust and thrust minor,
obtained in a fully automated way with the program CAESAR.  The
program recognises first that the two observables belong to the class
for which a NLL resummation is feasible and automatically determines
the input needed by the master formula.  It then exploits the master
formula to produce resummed curves such as the ones shown in
fig.~\ref{fig:T-Tmin}. These plots show the resummed differential
distributions (without matching with fixed order) for $\tau_{t,g}$ and
$T_{m,g}$ at the Tevatron II centre-of-mass energy $\sqrt s = 1.96
\mbox{TeV}$. The dijet events are selected by requiring two hard jets
with $E_t> 50 \mbox{GeV}$ and $|\eta|<1$. We use the CTEQ6M parton
distributions \cite{CTEQ} corresponding to $\as(M_Z)=0.118$ and set
both the renormalisation and facorisation scale at the partonic
centre-of-mass energy.  From the two distributions clearly emerges the
 separation among the various partonic channels, information that
can be exploited for fits of parton distributions.
\vspace*{-1cm}

\section{Conclusions and outlook}
The study of event shapes and jet rates in hadron hadron collisions is
particularly important for the understanding of QCD dynamics.  We
have now a computer code that in a fully automated way provides the
resummed distribution of any suitable jet observable in an arbitrary hard
process. Much work remains still to be done, both to refine the
existing code and to include  automated matching with fixed
order results. Nevertheless we believe that such a program will
open the way to a vast amount of phenomenological studies.

\section*{Acknowledgements} 
I thank the organisers for the possibility to participate to DIS2003
in an enchanting town such as St. Petersburg. I am particularly
grateful to the organisers of the hadronic final state session, in
particular Yuri Dokshitzer, for the friendly and stimulating
atmosphere during the conference.

\end{document}